\documentclass[a4j]{article}

\usepackage{amsmath}
\usepackage{amssymb}
\begin{document}
\begin{flushright}
\end{flushright}
\vspace{44pt}
\begin{center}
\begin{Large}
\bf
$U_q(\widehat{sl_n})$-analog of
the XXZ chain with a boundary.
\end{Large}\\[6pt]
\vspace{30pt}
H. Furutsu ~and~ T. Kojima
\vspace{25pt}

{\it Department of Mathematics\\
College of Science and Technology,
Nihon University,\\
 1-8,Kanda-Surugadai,
Chiyoda Tokyo 101, JAPAN}

\vspace{60pt}
\underline{Abstract}
\end{center}
We study $U_q(\widehat{sl_n})$ analog of the XXZ spin chain
with a boundary magnetic field $h$.
We construct explicit bosonic formulas of the vacuum vector
and the dual vacuum vector with a boundary magnetic field.
We derive integral formulas
of the correlation functions.
\vspace{15pt}

\newpage
\section{Introduction}
In the standard treatment
of quantum integrable systems,
one starts with a finite box and impose periodic 
boundary conditions,
in order to ensure integrability.
Recently, there has been increasing interest in exploring other 
possible boundary conditions
compatible with integrability.

For the free fermionic models,
there have been obtained 
many explicit formulas of the correlation functions
with non-periodic boundary conditions.
In this category,
the work on the two dimensional Ising model by 
B.M. McCoy and T.T. Wu
\cite{MW} are among the earliset.
They derived the spin-spin correlation functions
with a boundary field, by combinatorial arguments.
For an impenetrable Bose gas model,
T. Kojima derived 
the ground state correlation functions \cite{K1}
and the time dependent correlation functions 
\cite{K2}
with Dirichlet or Neumann conditions.

In this paper we are interested in the non free fermion 
model.
For the non free fermion model,
E.K. Sklyanin 
\cite{S} began a systematic approach 
to open boundary
problem, so-called open boundary Bethe Ansatz.
He formulated the transfer matrix to open boundary problem, 
and derived the Bethe Ansatz equations.
Jimbo et al.\cite{JKKKM}
 united Sklyanin's open boundary Bethe Ansatz
and Kyoto school's method \cite{JM}
- so called 
representation  theory approach to solvable models.
They studied XXZ model with a boundary,which are governed
by the quantum affine symmmetry $U_q(\widehat{sl_2})$.
They constructed 
explicit bosonic formulas of the vacuum vector
with an arbitrary boundary magnetic field.
They derived integral formulas
of the boundary magnetizations.
H. Ozaki \cite{O} studied 
$U_q(\widehat{sl_3})$ analog of
the XXZ chains with a boundary.
He constructed explicit bosonic formulas of 
the vacuum vector
for the special boundary conditions.
In this paper we studied $U_q(\widehat{sl_n})$
analog of the XXZ chains with 
an arbitrary boundary magnetic field $h$.
Our results are new even for $U_q(\widehat{sl_3})$ case.

The Hamiltonian of our model is given by
\begin{eqnarray}
{\it H_B}=
\sum_{k=1}^\infty
\left\{
\cosh(\gamma)
\sum_{a,b=0}^{n-1}e_{aa}^{(k+1)}e_{bb}^{(k)}
+
\sinh(\gamma)\sum_{a,b=0}^{n-1}
{\rm sgn(b-a)}e_{aa}^{(k+1)}e_{bb}^{(k)}
\right.\nonumber\\
~~~~~~~~~~
\left.-\sum_{a,b=0\atop{a \neq b}}
^{n-1}e_{ab}^{(k+1)}e_{ba}^{(k)} 
\right\}
+h\sum_{a=L}^{M-1}e_{aa}^{(1)}-
2\sinh(\gamma)\left\{\sum_{a=0}^{L-1}e_{aa}^{(1)}-
\sum_{a=M}^{n-1}
e_{aa}^{(1)}\right\},\label{Hamiltonian}
\end{eqnarray}
where
$0\leq \gamma <+\infty$
and $0\leq L \leq M \leq n-1$.
The Hamiltonian acts on the semi infinite tensor
products
of ${\mathbb C}^n$.
We construct 
explicit bosonic formulas of the vacuum vector
and the dual vacuum vector with a boundary magnetic field.
Using bosonization of the vacuum, the
dual vacuum and the vertex operators
\cite{Koy},
we derive
integral formulas of the correlation functions.

Now a few words about the organization of this paper.
In Section 2, we formulate our problem.
In Section 3, we construct explicit bosonic formulas of 
the vacuum vector and the dual vacuum vector.
In Section 4, we derive 
integral formulas of the correlation functions.
In Appendix we summarize the bosonizations
of the Vertex operators, for reader's convenience.

\section{Formulation}
The purpose of this section is to
formulate our problem.
\subsection{Notation}
We fix a real number $-1<q<0$ and 
an integer $n \in {\mathbb N}-\{0,1\}$.
In the sequel, we denote 
$(q^k-q^{-k})/(q-q^{-1})$
by $[k]$.
Let $P$ be a free Abelian group on letters 
$\Lambda_1,\cdots,\Lambda_{n-1},~\delta$.
$$
P=\oplus_{i=0}^{n-1}{\mathbb Z}\Lambda_i \oplus 
{\mathbb Z}\delta.
$$
We call $P$ the weight lattice.
Let $h_1,\cdots,h_{n-1},d$ be an ordered basis of $P^*=
{\rm Hom}(P, {\mathbb Z})$ dual to 
$\Lambda_1,\cdots,\Lambda_{n-1},~\delta$.
$$
\langle \Lambda_i,h_j \rangle=\delta_{ij},~
\langle \Lambda_i,d\rangle=0,~
\langle \delta, h_j \rangle=0,~
\langle \delta, d \rangle=1.
$$
Let us set the simple roots as
$$
\alpha_0=-\Lambda_{n-1}+2\Lambda_0-\Lambda_1+\delta,~
\alpha_j=-\Lambda_{j-1}+
2\Lambda_j-\Lambda_{j+1},~(j=1,\cdots,n-1).
$$
The projection to classical lattice is given by
$$
\bar{\Lambda}_i=\Lambda_i-\Lambda_0,~~\bar{\delta}=0.
$$
The invariant bilinear form on 
$(\cdot|\cdot):P \times P \to {\mathbb Z}$
by
$$
(\alpha_i|\alpha_j)=-\delta_{i,j-1}+2\delta_{i,j}
-\delta_{i,j+1},~~(\delta|\delta)=0.
$$
The quantum affine algebras $U_q(\widehat{sl_n})$
are algebras with $1$ over ${\mathbb C}$, defined by 
the generators $e_i, f_i, t_i^{\pm 1}=q^{\pm h_i}, 
q^d, (i=0, \cdots, n-1)$
through the following defining relations :
\begin{eqnarray*}
t_it_j=t_jt_i,~t_ie_jt_i^{-1}=q^{\langle \alpha_j,h_i\rangle}
e_j,~t_if_jt_i^{-1}=q^{-\langle \alpha_j,h_i\rangle}f_j,
\end{eqnarray*}
\begin{eqnarray*}
\left[ e_i,f_j \right]=\delta_{i,j}\frac{t_i-t_i^{-1}}
{q-q^{-1}},
\end{eqnarray*}
\begin{eqnarray*}
\sum_{k=0}^b(-1)^k\left[\begin{array}{c}b\\k
\end{array}\right]e_i^k e_j e_i^{b-k}=0,~
\sum_{k=0}^b(-1)^k\left[\begin{array}{c}b\\k
\end{array}\right]f_i^k f_j f_i^{b-k}=0,
\end{eqnarray*}
where we have set
\begin{eqnarray*}
b=1-\langle \alpha_i, h_j\rangle,
\left[\begin{array}{c}b\\k
\end{array}\right]=\frac{[b]!}{[k]![b-k]!},~
[k]!=[k][k-1]\cdots[1].
\end{eqnarray*}
Let us set $U_q'(\widehat{sl_n})$ 
be the subalgebra of $U_q(\widehat{sl_n})$ generated by
$t_i,e_i,f_i,(i=0,\cdots,n-1)$.
Let us set $U_q(sl_n)$ be the subalgebra of
 $U_q(\widehat{sl_n})$ generated by
$t_i,e_i,f_i,(i=1,\cdots,n-1)$. 
We denote the irreducible highest weight $U_q$-module
with highest weight $\lambda$ by $V(\lambda)$.
Let $V$ be a finite dimensional representation of
$U_q(sl_n)$. The evalutation module $V_z=V\otimes {\mathbb C}
[z,z^{-1}]$ in the homogeneous picture is the following
$U_q(\widehat{sl_n})$-module defined by
$$
e_0(v \otimes z^m)=(f_1 v)\otimes z^{m+1},
e_j(v \otimes z^m)=(e_j v)\otimes z^{m},~(j=1,\cdots,n),
$$
$$
f_0(v \otimes z^m)=(e_1 v)\otimes z^{m-1},
f_j(v \otimes z^m)=(f_j v)\otimes z^{m},~(j=1,\cdots,n),
$$
$$t_0=t_1^{-1},~t_1(v \otimes z^m)=(t_1v) \otimes z^m,
d=z\frac{d}{dz}.
$$
\subsection{Solvable Model}
Fix the number $i \in \{0,1,\cdots,n-1\}$.
Let 
$V={\mathbb C}v_0 \oplus {\mathbb C}v_1 \oplus
\cdots \oplus {\mathbb C}v_{n-1}$ 
be a basic representation of $U_q(sl_n)$.
Let
the R-matrix $R^{(i)}
(z_1/z_2) \in {\rm End}(V_{z_1}\otimes V_{z_2})$
be an intertwiner of $U_q(\widehat{sl_n})$ 
in the homogeneous picture,
$$
PR^{(i)}(z_1/z_2) : V_{z_1}\otimes V_{z_2} \rightarrow
 V_{z_2}\otimes V_{z_1}. 
$$ 
Fixing the normalization constant, the R-matrix $R(z)$
is given by
\begin{eqnarray*}
R^{(i)}(z)_{m,j}^{k,l}&=&\frac{1}{\kappa^{(i)}(z)}\times
\left\{
\begin{array}{cc}
\displaystyle
\frac{(1-q^2)\sqrt{z}}{1-q^2z}\delta_{m,k}\delta_{j,l}
\sqrt{z}^{~{\rm sgn}(k-l)}
+\frac{(1-z)q}{1-q^2z}\delta_{m,l}\delta_{j,k},&
0\leq k\neq l \leq n-1,\\
\delta_{m,k},&0 \leq k=l \leq n-1.
\end{array}\right.
\end{eqnarray*}
Here we have set
$$
\kappa^{(i)}(z) = z^{\delta_{i,0}}
\frac{(q^2z^{-1};q^{2n})_\infty
(q^{2n}z ;q^{2n})_\infty}
{(q^2z;q^{2n})_\infty(q^{2n}z^{-1};q^{2n})_\infty},
$$
where $(z;p)_\infty = \prod_{n=1}^\infty (1-zp^n)$ .\\
The R-matrix $R^{(i)}(z)$ satisfies the Yang-Baxter equation.
Let us fix the integer number $0\leq L \leq M \leq n-1$ and
$r \in {\mathbb R}$.
Let us set the reflection K-matrix $K^{(i)}(z) 
\in {\rm End}(V_z)$
\cite{O}, \cite{VR}, 
by
\begin{eqnarray}
K^{(i)}(z)=\frac{\varphi^{(i)}(z)}
{\varphi^{(i)}(1/z)}\left(
\begin{array}{ccccc}
\frac{k_0(z)}{k_0(1/z)}& & & & \\
& \frac{k_1(z)}{k_1(1/z)} & & & \\
& &\cdots & & \\
& & & \frac{k_{n-2}(z)}{k_{n-2}(1/z)}& \\ 
& & & & \frac{k_{n-1}(z)}{k_{n-1}(1/z)}
\end{array}\right),\nonumber
\end{eqnarray}
where we have set
$$
k_{0}(z)=\cdots=k_{L-1}(z)=z,
$$
$$
k_{L}(z)=\cdots=k_{M-1}(z)=1-rz,
$$
$$
k_{M}(z)=\cdots=k_{n-2}(z)=k_{n-1}(z)=1.
$$
The scalar functions
$\varphi^{(i)}(z)$ are given by 
(\ref{scalar1}), (\ref{scalar2}), 
(\ref{scalar3}), (\ref{scalar4}),
(\ref{scalar5}), (\ref{scalar6}),
and (\ref{scalar7}).
The refrection matrix $K^{(i)}(z)$ satisfies
the Boundary Yang-Baxter equation.
$$
K_2^{(i)}(z_2)R_{21}^{(i)}(z_1z_2)
K_1^{(i)}(z_1)R_{12}^{(i)}(z_1/z_2)=
R_{21}^{(i)}(z_1/z_2)K_1^{(i)}(z_1)
R_{12}^{(i)}(z_1z_2)K_2^{(i)}(z_2).
$$
~\\
{\it Note.~
de Vega and 
Ruiz \cite{VR} found the special diagonal 
solutions for the case $L=0, 1\leq M \leq n-1$
and $0\leq L=M \leq n-1$.
Ozaki \cite{O} found the general diagonal solutions
as the same arguments as in \cite{VR}.}

~\\

Graphically, an elements of
the R-matrix 
$R^{(i)}(z)_{mj}^{kl}$
is
the picture described in Figure 1.
An element of the reflection K-matrix
$K^{(i)}(z)_j^k$ is the picture described
in Figure 2.\\
~\\

\unitlength 0.1in
\begin{picture}(39.70,16.95)(4.30,-17.45)
%
\special{pn 8}%
\special{pa 1417 369}%
\special{pa 1417 1339}%
\special{fp}%
\special{sh 1}%
\special{pa 1417 1339}%
\special{pa 1437 1272}%
\special{pa 1417 1286}%
\special{pa 1397 1272}%
\special{pa 1417 1339}%
\special{fp}%
\special{pa 1902 854}%
\special{pa 931 854}%
\special{fp}%
\special{sh 1}%
\special{pa 931 854}%
\special{fp}%
\special{sh 1}%
\special{pa 931 854}%
\special{pa 998 874}%
\special{pa 984 854}%
\special{pa 998 834}%
\special{pa 931 854}%
\special{fp}%
%
\special{pn 8}%
\special{pa 2872 514}%
\special{pa 4165 837}%
\special{fp}%
\special{sh 1}%
\special{pa 4165 837}%
\special{pa 4105 801}%
\special{pa 4113 824}%
\special{pa 4095 840}%
\special{pa 4165 837}%
\special{fp}%
\special{pa 4165 837}%
\special{pa 2863 1128}%
\special{sh 1}%
\special{pa 2863 1128}%
\special{pa 2932 1133}%
\special{pa 2915 1116}%
\special{pa 2924 1094}%
\special{pa 2863 1128}%
\special{fp}%
%
\special{pn 8}%
\special{pa 4157 190}%
\special{pa 4157 1484}%
\special{fp}%
\put(31.3000,-3.6100){\makebox(0,0){$k$}}%
\put(31.3000,-13.2200){\makebox(0,0){$j$}}%
\put(28.0700,-3.5200){\makebox(0,0){$z$}}%
\put(28.0700,-13.2200){\makebox(0,0){$z^{-1}$}}%
\put(15.7800,-6.8400){\makebox(0,0){$z$}}%
\put(7.7000,-8.4600){\makebox(0,0){$j$}}%
\put(14.2400,-2.0700){\makebox(0,0){$k$}}%
\put(20.4700,-8.5400){\makebox(0,0){$l$}}%
\put(14.2000,-18.3000){\makebox(0,0){\bf Figure 1}}%
\put(34.2000,-18.3000){\makebox(0,0){\bf Figure 2}}%
%
\special{pn 8}%
\special{pa 4400 220}%
\special{pa 4400 1420}%
\special{da 0.070}%
\special{pa 4270 210}%
\special{pa 4270 1410}%
\special{da 0.070}%
\special{pa 4220 190}%
\special{pa 4220 1390}%
\special{da 0.070}%
\special{pa 4340 210}%
\special{pa 4340 1410}%
\special{da 0.070}%
\special{pa 430 50}%
\special{pa 430 90}%
\special{da 0.070}%
\put(14.1000,-15.2000){\makebox(0,0){$m$}}%
\end{picture}%
\\~\\
The Type-I Vertex operator 
$\Phi_{\lambda}^{\mu,V}(z)$ is an intertwining
operator of $U_q(\widehat{sl_n})$ defined by
$$
\Phi_{\lambda}^{\mu,V}(z):~
V(\lambda)\rightarrow V(\mu)\hat{\otimes}V_z.
$$
The Type-I dual Vertex operator is an intertwing
operator of $U_q(\widehat{sl_n})$ defined by
$$
\Phi_{\mu,V}^{\lambda}(z):~
V(\mu) \otimes V_z \rightarrow \hat{V}(\lambda).
$$ 
Lut us set the components of the vertex operators 
$\Phi_{\lambda,j}^{\mu,V}(z)$
as follows.
$$
\Phi_{\lambda}^{\mu,V}(z)|u\rangle
=\sum_{j=1}^{n-1}\Phi_{\lambda,j}^{\mu,V}(z)
|u\rangle \otimes v_j,~~{\rm for}~~|u\rangle \in
V(\lambda).
$$
Lut us set the components of the vertex operators 
$\Phi_{\mu,V,j}^{\lambda}(z)$
as follows.
$$
\Phi_{\mu,V}^{\lambda}(z)(|u\rangle \otimes v_j)
=\Phi_{\mu,V,j}^{\lambda}(z)|u\rangle,
~~{\rm for}~~|u\rangle \in
V(\mu).
$$
Only $\Phi_{V(\Lambda_{j+1})}^{V(\Lambda_j),V}(z),~
\Phi_{V(\Lambda_{j}),V}^{V(\Lambda_{j+1})}(z)$
are nontrivial. We take the following normalizations.
$$
\Phi_{V(\Lambda_{i+1})}^{V(\Lambda_i),V}(z)|\Lambda_{i+1}\rangle
=|\Lambda_i\rangle \otimes v_i+\cdots,
$$
$$
\Phi_{V(\Lambda_{i}),V}^{V(\Lambda_{i+1})}(z)
|\Lambda_{i}\rangle
\otimes v_i
=|\Lambda_{i+1} \rangle +\cdots,
$$
where $|\Lambda_i\rangle$ is the highest vector of $V(\Lambda_i)$.
Let us consider the product
of the vertex operators.
Let us fix the following notation.
\begin{eqnarray*}
&&
~^{\bf (1)}\Phi_{V(\Lambda_{i-1})}^{ V(\Lambda_{i-2}),V}(z_1)
~^{\bf (2)}\Phi_{V(\Lambda_{i})}^{ V(\Lambda_{i-1}),V}(z_2)\\
&=&
\sum_{j_1,j_2=0}^{n-1}
\Phi_{V(\Lambda_{i-1}),j_1}^{ V(\Lambda_{i-2}),V}(z_1)
\Phi_{V(\Lambda_{i}),j_2}^{ V(\Lambda_{i-1}),V}(z_2)
\otimes v_{j_1} \otimes v_{j_2}.
\end{eqnarray*}
The vertex operators satisfy
\begin{eqnarray*}
&&R^{(i)}(z_1/z_2)
~^{\bf (1)}\Phi_{V(\Lambda_{i-1})}^{ V(\Lambda_{i-2}),V}(z_1)
~^{\bf (2)}\Phi_{V(\Lambda_{i})}^{ V(\Lambda_{i-1}),V}(z_2)\\
&&=
~^{\bf (2)}\Phi_{V(\Lambda_{i-1})}^{ V(\Lambda_{i-2}),V}(z_2)
~^{\bf (1)}\Phi_{V(\Lambda_{i})}^{ V(\Lambda_{i-1}),V}(z_1).
\end{eqnarray*}
In the sequel, we use the abberivations
$$
\Phi_{j}^{(i,i+1)}(z)=
\Phi_{V(\Lambda_{i+1}),j}^{V(\Lambda_i),V}(z),~~~
\Phi_{j}^{*(i+1,i)}(z)=
\Phi_{V(\Lambda_{i}),V,j}^{V(\Lambda_{i+1})}(z)
$$
Graphically, the vertex operator is the picture 
described in Figure 3.
The dual vertex operator is the picture 
described in Figure 4. 

~\\
\unitlength 0.1in
\begin{picture}(30.95,16.10)(-4.30,-17.25)
%
\special{pn 8}%
\special{pa 600 400}%
\special{pa 2600 400}%
\special{fp}%
\special{sh 1}%
\special{pa 2600 400}%
\special{pa 2533 380}%
\special{pa 2547 400}%
\special{pa 2533 420}%
\special{pa 2600 400}%
\special{fp}%
\special{pa 2200 200}%
\special{pa 2200 600}%
\special{fp}%
\special{sh 1}%
\special{pa 2200 600}%
\special{pa 2220 533}%
\special{pa 2200 547}%
\special{pa 2180 533}%
\special{pa 2200 600}%
\special{fp}%
\special{pa 2000 200}%
\special{pa 2000 600}%
\special{fp}%
\special{sh 1}%
\special{pa 2000 600}%
\special{pa 2020 533}%
\special{pa 2000 547}%
\special{pa 1980 533}%
\special{pa 2000 600}%
\special{fp}%
\special{pa 1800 200}%
\special{pa 1800 600}%
\special{fp}%
\special{sh 1}%
\special{pa 1800 600}%
\special{pa 1820 533}%
\special{pa 1800 547}%
\special{pa 1780 533}%
\special{pa 1800 600}%
\special{fp}%
\special{pa 2600 1200}%
\special{pa 600 1200}%
\special{fp}%
\special{sh 1}%
\special{pa 600 1200}%
\special{pa 667 1220}%
\special{pa 653 1200}%
\special{pa 667 1180}%
\special{pa 600 1200}%
\special{fp}%
\special{pa 2200 1000}%
\special{pa 2200 1400}%
\special{fp}%
\special{sh 1}%
\special{pa 2200 1400}%
\special{pa 2220 1333}%
\special{pa 2200 1347}%
\special{pa 2180 1333}%
\special{pa 2200 1400}%
\special{fp}%
\special{pa 2000 1000}%
\special{pa 2000 1400}%
\special{fp}%
\special{sh 1}%
\special{pa 2000 1400}%
\special{pa 2020 1333}%
\special{pa 2000 1347}%
\special{pa 1980 1333}%
\special{pa 2000 1400}%
\special{fp}%
\special{pa 1800 1000}%
\special{pa 1800 1400}%
\special{fp}%
\special{sh 1}%
\special{pa 1800 1400}%
\special{pa 1820 1333}%
\special{pa 1800 1347}%
\special{pa 1780 1333}%
\special{pa 1800 1400}%
\special{fp}%
\put(2.0000,-4.0000){\makebox(0,0){$\Phi_j(z)=$}}%
\put(10.0000,-2.0000){\makebox(0,0){$z$}}%
\put(28.0000,-4.1000){\makebox(0,0){$j$}}%
\put(28.0000,-12.1000){\makebox(0,0){$j$}}%
\put(2.0000,-12.1000){\makebox(0,0){$\Phi^*_j(z)=$}}%
\put(10.0000,-14.2000){\makebox(0,0){$z^{-1}$}}%
\put(13.9000,-8.0000){\makebox(0,0){\bf Figure 3}}%
\put(14.0000,-18.1000){\makebox(0,0){\bf Figure 4}}%
\end{picture}%

~\\
~\\
We define the normalized transfer matrix by
\begin{eqnarray}
T_B^{(i)}(z)=g_n \sum_{j=0}^{n-1}
\Phi_j^{*(i,i-1)}(z^{-1})K^{(i)}(z)_j^j
\Phi_j^{(i-1,i)}(z),
\end{eqnarray}
where we have used
$$
g_n=\frac{(q^2;q^{2n})_\infty}
{(q^{2n};q^{2n})_\infty}.
$$
Graphically, the transfer matrix $T_B^{(i)}(z)$
in the semi-infinite chain,
is the picture in Figure 5.
It describes a semi-infinite two-dimensional
lattice, with alternating spectral parameter.
\\~\\
\unitlength 0.1in
\begin{picture}(47.85,15.90)(-4.75,-17.05)
%
\special{pn 8}%
\special{pa 400 390}%
\special{pa 4000 790}%
\special{fp}%
\special{sh 1}%
\special{pa 4000 790}%
\special{pa 3936 763}%
\special{pa 3947 784}%
\special{pa 3932 803}%
\special{pa 4000 790}%
\special{fp}%
\special{pa 4000 790}%
\special{pa 380 1050}%
\special{fp}%
\special{sh 1}%
\special{pa 380 1050}%
\special{pa 448 1065}%
\special{pa 433 1046}%
\special{pa 445 1025}%
\special{pa 380 1050}%
\special{fp}%
\special{pa 3410 220}%
\special{pa 3410 1420}%
\special{fp}%
\special{sh 1}%
\special{pa 3410 1420}%
\special{pa 3430 1353}%
\special{pa 3410 1367}%
\special{pa 3390 1353}%
\special{pa 3410 1420}%
\special{fp}%
\special{pa 2990 210}%
\special{pa 2990 1410}%
\special{fp}%
\special{sh 1}%
\special{pa 2990 1410}%
\special{pa 3010 1343}%
\special{pa 2990 1357}%
\special{pa 2970 1343}%
\special{pa 2990 1410}%
\special{fp}%
\special{pa 2570 200}%
\special{pa 2570 1400}%
\special{fp}%
\special{sh 1}%
\special{pa 2570 1400}%
\special{pa 2590 1333}%
\special{pa 2570 1347}%
\special{pa 2550 1333}%
\special{pa 2570 1400}%
\special{fp}%
\special{pa 2210 210}%
\special{pa 2210 1410}%
\special{fp}%
\special{sh 1}%
\special{pa 2210 1410}%
\special{pa 2230 1343}%
\special{pa 2210 1357}%
\special{pa 2190 1343}%
\special{pa 2210 1410}%
\special{fp}%
\special{pa 1780 190}%
\special{pa 1780 1390}%
\special{fp}%
\special{sh 1}%
\special{pa 1780 1390}%
\special{pa 1800 1323}%
\special{pa 1780 1337}%
\special{pa 1760 1323}%
\special{pa 1780 1390}%
\special{fp}%
\special{pa 1370 200}%
\special{pa 1370 1400}%
\special{fp}%
\special{sh 1}%
\special{pa 1370 1400}%
\special{pa 1390 1333}%
\special{pa 1370 1347}%
\special{pa 1350 1333}%
\special{pa 1370 1400}%
\special{fp}%
\special{pa 1020 210}%
\special{pa 1020 1410}%
\special{fp}%
\special{sh 1}%
\special{pa 1020 1410}%
\special{pa 1040 1343}%
\special{pa 1020 1357}%
\special{pa 1000 1343}%
\special{pa 1020 1410}%
\special{fp}%
\put(5.2000,-2.0000){\makebox(0,0){$z$}}%
\put(4.7000,-13.2000){\makebox(0,0){$z^{-1}$}}%
\put(22.0000,-17.9000){\makebox(0,0){\bf Figure 5}}%
\put(2.0000,-7.9000){\makebox(0,0){$T_B^{(i)}(z)=$}}%
%
\special{pn 8}%
\special{pa 3990 200}%
\special{pa 3990 1400}%
\special{fp}%
%
\special{pn 8}%
\special{pa 4100 210}%
\special{pa 4100 1410}%
\special{da 0.070}%
\special{pa 4210 210}%
\special{pa 4210 1410}%
\special{da 0.070}%
\special{pa 4310 210}%
\special{pa 4310 1410}%
\special{da 0.070}%
\end{picture}%

~\\
The renormalized Hamiltonian $H_B^{(i)}$
in (\ref{Hamiltonian})
is then defined by
\begin{eqnarray}
\left.\frac{d}{dz}T_B^{(i)}(z)\right|_{z=1}=
\frac{q}{1-q^2}H_B^{(i)}+const,
\end{eqnarray}
where we set
$$h=\frac{1-q^2}{q}\times \frac{r+1}{r-1},~~~
q=-e^{-\gamma}.$$
Here the right hand side
$H_B^{(i)}$ acts on the space ${\mathcal H}^{(i)}$,
where ${\mathcal H}^{(i)}$ is the span of vectors
$|p\rangle = \otimes_{k=1}^\infty v_{p(k)}$,
called paths, labelled by maps
$p: {\mathbb Z}\geq 1 \to {\mathbb Z}/n {\mathbb Z}$
satisfying the asymptotic boundary condition
$$
p(k)=k+i \in 
\left\{0,1,2,\cdots, n-1
\right\}={\mathbb Z}/n {\mathbb Z},
~~{\rm for}~k>>1. 
$$
We have identified the highest weight module $V(\Lambda_i)$
and the path space ${\mathcal H}^{(i)}$, 
following the strategy proposed in
Ref. \cite{JM}.
In order to diagonalize the Hamiltonian $H_B^{(i)}$
in (\ref{Hamiltonian}),
we diagonalize the transfer matrix $T_B^{(i)}(z)$.
Using the Boundary Yang-Baxter equations,
we have
$$
[T_B^{(i)}(z),T_B^{(i)}(z')]=0,~~
T_B^{(i)}(1)=id,~~
T_B^{(i)}(z)T_B^{(i)}(z^{-1})=id.
$$
This commuting relation of the transfer matrix asserts
the integrability of this problem.

\section{Vacuum Vectors}
The purpose of this section is to construct
the explicit bosonic formulas of the vacuum vector
$|i\rangle_B$ such that
\begin{eqnarray}
T_B^{(i)}(z)|i\rangle_B=|i\rangle_B,~~(i=0,\cdots , n-1),
\label{t-eigen1}
\end{eqnarray}
which is realized as
$$
|i\rangle_B=e^{F_i}|i\rangle,
$$
where $|i\rangle$ is the highest weight vector of $V(\Lambda_i)$,
and $F_i$ is a quadratic in the boson operators.
Multiplying the vertex operator $\Phi^{(i-1,i)}_j(z^{-1})$
from the left, and using the inversion relation,
$$
g_n \Phi^{(i-1,i)}_j(z)\Phi^{*~(i,i-1)}_j(z)=id,~~
g_n=\frac{(q^2;q^{2n})_\infty}{(q^{2n};q^{2n})_\infty},
$$
we know the eigenvalue problem (\ref{t-eigen1}) is equivalent
to
\begin{eqnarray}
K^{(i)}(z)_j^j\Phi^{(i-1,i)}_j(z)|i\rangle_B=
\Phi^{(i-1,i)}_j(z^{-1})|i\rangle_B.\label{basic1}
\end{eqnarray}
We construct the dual 
vacuum vectors $~_B\langle i |$ such that
\begin{eqnarray}
~_B\langle i | T_B^{(i)}(z)
=~_B\langle i |,~~~
(i=0, \cdots , n-1),\label{t-eigen2}
\end{eqnarray}
which is realized as
$$
~_B\langle i |
=\langle i |e^{G_i},
$$
where $\langle i|$ is the lowest weight vector of 
the restricted dual module $V^*(\Lambda_i)$,
and $G_i$ is a quadratic in the boson operators.
As the same argument as the vacuum vectors,
we know the eigenvalue problem (\ref{t-eigen2})
is equivalent to
\begin{eqnarray}
K^{(i)}(z)_j^j~_B\langle i |\Phi^{*~(i,i-1)}_j(z^{-1})=
~_B\langle i |\Phi^{*~(i,i-1)}_j(z).\label{basic2}
\end{eqnarray}
The scalar factor of the refrection matrix 
$\varphi^{(i)}(z)$ are given by
(\ref{scalar1}), (\ref{scalar2}), (\ref{scalar3}),
(\ref{scalar4}), (\ref{scalar5}), (\ref{scalar6}),
and (\ref{scalar7}).

~\\
{\it Note.~
H. Ozaki \cite{O} constructed the vacuum vector 
$|0\rangle_B$ for
$U_q(\widehat{sl_3}),~V(\Lambda_0)$, 
$(a)~L=M=2$ or $(b)~L=0, M=2$ cases.
Our results are new for $U_q(\widehat{sl_3})$-case.} 
\subsection{Vacuum}
Let us consider
the vacuum vector $|i\rangle_B$.
Since the total spin is conserved,
it should be a linear combination of the states
created by the oscillators $a_s(-k)$ over the highest
weight vector $|i\rangle$.
We make the ansatz that it has the following form.
$$
|i\rangle_B=e^{F_i}|i\rangle,
$$
where
$$
F_i=\sum_{s,t=1}^{n-1}
\sum_{k=1}^\infty
\alpha_{s,t}(k)a_s(-k)a_t(-k)+
\sum_{s=1}^{n-1}\sum_{k=1}^\infty
\beta_s^{(i)}(k)a_s(-k).
$$
The operator $e^{F_i}$ has the effect of a Bogoliubov
transformation,
\begin{eqnarray*}
e^{-F_i}a_j(k)e^{F_i}=a_j(k)&+&\sum_{s,t=1}^{n-1}
\alpha_{s,t}(k)\left(\frac{[(a_j|a_s)k][k]}{k}a_t(-k)
+(s \leftrightarrow t)\right)\\
&+&\sum_{s=1}^{n-1}\beta_s^{(i)}(k)\frac{[(a_j|a_s)k][k]}{k}.
\end{eqnarray*}
Using the bosonic formulas of
the vertex operators, 
we have 
the $(n-1)$-th component of the equation (\ref{basic1})
as follows,
$$
\varphi^{(i+1)}(z)z^{\frac{n-i-1}{n}+(\bar{\Lambda}_{n-1}|
\bar{\Lambda}_{i+1})}
e^{P(z)}q^{Q(z)}e^{F_{i+1}}|i+1\rangle=(z \leftrightarrow z^{-1}).
$$
Comparing the bosonic parts of the both sides,
we have
$$
\alpha_{s,n-1}(k)=-\frac{1}{2}
\frac{[sk]k}{[k]^2[nk]}q^{(2n+2)k},~~(1\leq s \leq n-1).
$$
Comparing the bosonic part of the 
$j$-th component of the equation
(\ref{basic1}), we have
\begin{eqnarray}
\alpha_{s,t}(k)=\frac{-kq^{2(n+1)k}}{2[k]}\times
I_{s,t}(k).
\label{result}
\end{eqnarray}
Here the matrix
$(I_{s,t}(k))_{1\leq s,t \leq n-1}$
is the inverse matrix of the A-type Cartan matrix
$([(a_s|a_t)k])_{1\leq s,t \leq n-1}$.
More explicitly
\begin{eqnarray}
I_{s,t}(k)=\frac{[sk][(n-t)k]}{[k]^2[nk]}=
I_{t,s}(k),~~(1\leq s \leq t \leq n-1).
\label{inverse}
\end{eqnarray}
Using the explicit formulas of $\alpha_{s,t}(k)$,
we have the simple formulas of
the action of
the basic operators to the vacuum vectors.
\begin{eqnarray*}
e^{Q(z)}|i\rangle_B&=&h^{(i)}(z)e^{P(1/z)}|i\rangle_B,\\
e^{S_j^-(w)}|i \rangle_B&=&g_j^{(i)}(w)
e^{R_j^-(q^{2(n+1)}/w)}|i \rangle_B,~(1\leq j \leq n-1),
\end{eqnarray*}
where
$$
h^{(i)}(z)=\exp\left(
-\frac{1}{2}\sum_{k=1}^\infty
\frac{[(n-1)k]}{[nk]k}q^k z^{-2k}
-\sum_{k=1}^\infty \frac{[k]}{k}
\beta_{n-1}^{(i)}(k)q^{-(2n+1)k/2}z^{-k}
\right),
$$
and
$$
g_j^{(i)}(w)=
\exp\left(
-\frac{1}{2}\sum_{k=1}^\infty
\frac{[2k]q^{(2n+3)k}}{k[k]}w^{-2k}
+\sum_{k=1}^\infty
\sum_{s=1}^{n-1}\beta_s^{(i)}(k)
\frac{[(a_s|a_j)k]}{k}q^{k/2}w^{-k}\right).
$$
The $(n-1)$-th component of the equation (\ref{basic1})
reduces to
\begin{eqnarray}
\varphi^{(i)}(z)=z^{\delta_{i,0}-1}h^{(i)}(z^{-1}),~~
(0\leq i \leq n-1).
\end{eqnarray}
When we find 
the functions $g_j^{(i)}(w),~(1\leq j \leq n-1)$,
we can determine
both $\beta^{(i)}_j(k),~(1 \leq j \leq n-1)$,
and $h^{(i)}(z)$.

First we consider the case $|0\rangle_B$, and
$0 \leq L <M \leq n-1$.
We show the following pair of $g_j^{(0)}(q^{n+1}w)$
give the vacuum vector.
\begin{eqnarray}
g_j^{(0)}(q^{n+1}w)=\left\{
\begin{array}{cc}
(1-1/w^2)(1-q^{-n+2M-L}/(rw)),& j=L\\
(1-1/w^2)(1-q^{n-M}r/w),& j=M \\
(1-1/w^2),& j\neq L,M.
\end{array}
\right.\label{g0}
\end{eqnarray}
The $(n-2)$-th component of the equation
(\ref{basic1}) reduces to
\begin{eqnarray*}
&&\oint \frac{dw_{n-1}}{2\pi i w_{n-1}}
\frac{z^{-1}k_{n-2}(z)w_{n-1}g_{n-1}^{(0)}(q^{n+1}w_{n-1})}
{(1-qw_{n-1}/z)(1-qz/w_{n-1})(1-q/(zw_{n-1}))}\\
&\times&
e^{P(z)+P(1/z)+R_{n-1}^-(q^{n+1}w_{n-1})
+R_{n-1}^-(q^{n+1}/w_{n-1})}|0\rangle_B=
(z \leftrightarrow z^{-1}),
\end{eqnarray*}
where the contour encircles
$w=0,qz^{\pm1}$ but not $q^{-1}z^{\pm 1}$.
Because the bosonic part of this equation
is invariant under the change of variable
$w_{n-1} \to w_{n-1}^{-1}$ and
$z \to z^{-1}$,
this equation reduces to
the following integrand relation.
$$
\frac{g_{n-2}^{(0)}(q^{n+1}w)}
{g_{n-2}^{(0)}(q^{n+1}/w)}=-w^{-2}
\frac{k_{n-2}(z)/z(1-qz/w)-k_{n-2}(1/z)z(1-q/(zw))}
{k_{n-2}(z)/z(1-qzw)-k_{n-2}(1/z)z(1-qw/z)}.
$$
Therefore we have
\begin{eqnarray*}
g_{n-2}^{(0)}(q^{n+1}w)=
\left\{\begin{array}{cc}
1-1/w^2 ,& {\rm for~~}k_{n-2}(z)=1\\
(1-1/w^2)(1-rq/w),& {\rm for~~}k_{n-2}(z)=1-rz.
\end{array}\right.
\end{eqnarray*}
The $(n-k)$-th component
of the equation (\ref{basic1}) reduces to
\begin{eqnarray*}
&&\oint \frac{dw_{n-1}}{2\pi i w_{n-1}}
\cdots \oint \frac{dw_{n-k+1}}{2\pi i w_{n-k+1}}
w_{n-k+1}g_{n-1}^{(0)}(q^{n+1}w_{n-1})
\cdots g_{n-k+1}^{(0)}(q^{n+1}w_{n-k+1})\\
&\times&
z^{-1} k_{n-k}(z)
\frac{(1-qzw_{n-1})(1-qw_{n-1}w_{n-2})
\cdots (1-qw_{n-k+2}w_{n-k+1})}
{D(z,w_{n-1})D(w_{n-1},w_{n-2}) \cdots 
D(w_{n-k+2},w_{n-k+1})}e^{P(z)+P(1/z)}\\
&\times&
e^{R_{n-1}^-(q^{n+1}w_{n-1})+
R_{n-1}^-(q^{n+1}/w_{n-1})+\cdots+
R_{n-k+1}^-(q^{n+1}w_{n-k+1})+
R_{n-k+1}^-(q^{n+1}/w_{n-k+1})}|0\rangle_B\\
&=&(z \leftrightarrow z^{-1}),
\end{eqnarray*}
where we have set
$$D(w_1,w_2)=(1-qw_1/w_2)(1-qw_2/w_1)
(1-qw_1w_2)(1-q/(w_1w_2)).$$
Here the contour of the integral 
$\oint \frac{dw_j}{2\pi i w_j}$ 
encircles 
$0$ and $qw_{j+1}^{\pm 1}$ but not $q^{-1}w_{j+1}^{\pm 1}$.
$(w_{n}=z)$
Because the bosonic part of this equation and
the function $D(w_j,w_{j+1})$ are invariant under 
the change of variables $w_j \to w_j^{-1}$ and 
$z \to z^{-1}$, 
this equation reduces to
the following integrand relations.
\begin{eqnarray*}
\frac{g_{j}^{(0)}(q^{n+1}w)}
{g_{j}^{(0)}(q^{n+1}/w)}
=
\left\{
\begin{array}{cc}
-w^{-2}, & j \neq L, M, \\
\displaystyle
-w^{-2}
\frac{(1-q^{-n+2M-L}/(rw))}
{(1-q^{-n+2M-L}w/r)}, & j=L,\\
\displaystyle
-w^{-2}
\frac{(1-q^{n-M}r/w)}
{(1-q^{n-M}rw)}, & j=M.
\end{array}\right.
\end{eqnarray*}
Therefore we have the relation 
(\ref{g0}).

As the same arguments as the above,
we can construct $g_j^{(0)}(q^{n+1}w)$ for 
the case $0 \leq L=M \leq n-1$.
We have
\begin{eqnarray}
g_j^{(0)}(q^{n+1}w)=
\left\{
\begin{array}{cc}
1-1/w^2,& ~~j\neq L,\\
1-1/w^4,& ~~j=L.
\end{array}\right.
\end{eqnarray}
Now we have solved the problem for $|0\rangle_B$ case.
The coefficients of the bosonic
operators
are given by (\ref{result})
and
\begin{eqnarray}
&&\beta^{(0)}_j(k)=
(q^{(n+3/2)k}-q^{(n+1/2)k})\theta_k
\sum_{s=1}^{n-1}\hat{I}_{j,s}(k)
\label{result0}
\\
&+&\left\{
\begin{array}{cc}
-\hat{I}_{j,L}(k)q^{(2M-L+1/2)k}r^{-k}
-\hat{I}_{j,M}(k)q^{(2n-M+1/2)k}r^k,&~~(0\leq L<M\leq n-1)\\
-2(-1)^{k/2}\theta_k
\hat{I}_{j,L}(k)q^{(n+1/2)k}
,&~~(0\leq L=M\leq n-1).
\end{array}\right.\nonumber
\end{eqnarray}
Here we have used 
the symmetric matrix $\hat{I}_{s,t}(k)$ defined by
\begin{eqnarray}
\hat{I}_{s,t}(k)=\left\{
\begin{array}{cc}
0,&~(st=0),\\
I_{s,t}(k),&~(1\leq s,t \leq n-1),
\end{array}\right.
\label{hatI}
\end{eqnarray}
where we have used
the inverse matrix of the A-type Cartan
matrix $I_{s,t}(k)$~ (\ref{inverse}).
We have used 
\begin{eqnarray*}
\theta_k=\left\{
\begin{array}{cc}
1,& {\rm for}~~k=even,\\
0,& {\rm for}~~k=odd.
\end{array}\right.
\end{eqnarray*}
Using the explicit formulas of $\beta^{(0)}_{n-1}(k)$,
we have the scalar factor of the refrection matrix.
\begin{eqnarray}
&&\varphi^{(0)}(z)=
\frac{(q^{2n+2}z^2;q^{4n})_\infty}
{(q^{4n}z^2;q^{4n})_\infty}
\label{scalar1}
\\
&\times&
\left\{\begin{array}{cc}
\displaystyle
\frac{(rq^{2n}z;q^{2n})_\infty 
(r^{-1}q^{2M}z;q^{2n})_\infty}
{(rq^{2n-2M}z;q^{2n})_\infty 
(r^{-1}q^{2M-2L}z;q^{2n})_\infty}
,&~~(0\leq L < M \leq n-1),\\
\displaystyle
\frac{(-q^{2(n+L)}z^2;q^{4n})_\infty}
{(-q^{2(n-L)}z^2;q^{4n})_\infty},
&~~(0\leq L=M \leq n-1).\end{array}\right.
\nonumber
\end{eqnarray}
As the same arguments as the above,
we can construct $|i\rangle_B,~(1\leq i \leq n-1)$.
For $|i\rangle_B$ case, the integrand functions
satisfy the following relations.
\begin{eqnarray*}
\frac{g^{(i)}_j(q^{n+1}w)}
{g^{(i)}_j(q^{n+1}/w)}=w^{2\delta_{i,j}}
\frac{g^{(0)}_j(q^{n+1}w)}
{g^{(0)}_j(q^{n+1}/w)}.
\end{eqnarray*}
We have
\begin{eqnarray*}
g_j^{(L)}(q^{n+1}w)=
\left\{
\begin{array}{cc}\displaystyle
\frac{(1-1/w^2)}{(1-rq^{n-2M+L}/w)},&~j=L,\\
(1-1/w^2)(1-q^{n-M}r/w),&~j=M,\\
(1-1/w^2),&~j\neq L,M,
\end{array}
\right.
\left(
\begin{array}{cc}
1\leq i \leq n-1\\
i=L\\
0\leq L<M\leq n-1
\end{array}\right),
\end{eqnarray*}
\begin{eqnarray*}
g_j^{(M)}(q^{n+1}w)=
\left\{
\begin{array}{cc}
(1-1/w^2)(1-q^{-n+2M-L}/(rw)),&~j=L,\\
\displaystyle
\frac{(1-1/w^2)}{(1-q^{M-n}/(rw))},&~j=M,\\
(1-1/w^2),&~j\neq L,M,
\end{array}
\right.\left(
\begin{array}{cc}
1\leq i \leq n-1\\
i=M\\
0\leq L<M\leq n-1
\end{array}\right)
,
\end{eqnarray*}
\begin{eqnarray*}
g_j^{(i)}(q^{n+1}w)=
\left\{
\begin{array}{cc}
(1-1/w^2),&~j\neq i,L,M,\\
(1-1/w^2)(1-q^{-n+2M-L}/(rw)),&~j=L,\\
(1-1/w^2)(1-q^{n-M}r/w),&~j=M,\\
\displaystyle
\frac{(1-1/w^2)}
{(1+1/w^2)},&~j=i,
\end{array}
\right.~~\left(
\begin{array}{cc}
1\leq i \leq n-1\\
i\neq L,M\\
0\leq L<M\leq n-1
\end{array}\right),
\end{eqnarray*}
and
\begin{eqnarray*}
g_j^{(i)}(q^{n+1}w)=
\left\{
\begin{array}{cc}
(1-1/w^4),&~j=L,\\
(1-1/w^2),&~j\neq i,L,\\
\displaystyle
\frac{(1-1/w^2)}
{(1+1/w^2)},&~j=i,
\end{array}
\right.~~\left(
\begin{array}{cc}
1\leq i \leq n-1\\
i\neq L,
0\leq L=M\leq n-1
\end{array}\right),
\end{eqnarray*}
\begin{eqnarray*}
g_j^{(i)}(q^{n+1}w)=(1-1/w^2),
~~\left(
\begin{array}{cc}
1\leq i \leq n-1\\
0\leq L=M=i\leq n-1
\end{array}\right).
\end{eqnarray*}
The coefficients of the bosonic operators are given by
\begin{eqnarray}
&&\beta_j^{(i)}(k)=(q^{(n+3/2)k}-q^{(n+1/2)k})\theta_k
\sum_{s=1}^{n-1}\hat{I}_{j,s}(k)
\nonumber
\\
&+&
\left\{\begin{array}{cc}
\hat{I}_{j,L}(k)q^{(2n-2M+L+1/2)k}r^k
-\hat{I}_{j,M}(k)q^{(2n-M+1/2)k}r^k,&(i=L)\\
-\hat{I}_{j,L}(k)q^{(2M-L+1/2)k}r^{-k}
+\hat{I}_{j,M}(k)q^{(M+1/2)k}r^{-k},&(i=M)\\
-\hat{I}_{j,L}(k)q^{(2M-L+1/2)k}r^{-k}
-\hat{I}_{j,M}(k)q^{(2n-M+1/2)k}r^k&\\
+2(-1)^{k/2}\theta_k \hat{I}_{j,i}(k)q^{(n+1/2)k},
&(i\neq L,M
),
\end{array}
\right.\\
&&
~~~~~~~~~~~~~~~
\left(\begin{array}{c}
1\leq i \leq n-1\\
0\leq L<M \leq n-1
\end{array}\right),\nonumber
\end{eqnarray}
and
\begin{eqnarray}
&&\beta_j^{(i)}(k)=(q^{(n+3/2)k}-q^{(n+1/2)k})\theta_k
\sum_{s=1}^{n-1}\hat{I}_{j,s}(k) \\
&+&2(-1)^{k/2}\theta_k q^{(n+1/2)k}
(-\hat{I}_{j,L}(k)+\hat{I}_{j,i}(k)),
\left(\begin{array}{c}
1\leq i \leq n-1\\
0\leq L=M \leq n-1
\end{array}\right),\nonumber
\end{eqnarray}
where we have used 
the matrix $\hat{I}_{s,t}(k)$
defined in (\ref{hatI}).
The scalar factor of the refrection
matrix is given by
\begin{eqnarray}
&&\varphi^{(i)}(z)=z^{-1}\frac{(q^{2n+2}z^2;q^{4n})_\infty}
{(q^{4n}z^2;q^{4n})_\infty}
\label{scalar2}\\
&\times&
\left\{
\begin{array}{cc}
\displaystyle
\frac{(q^{2n}rz;q^{2n})_\infty}
{(q^{2L-2M+2n}rz;q^{2n})_\infty},&~(i=L),\\
\displaystyle
\frac{(r^{-1}z;q^{2n})_\infty}
{(q^{2M-2L}r^{-1}z;q^{2n})_\infty},&~(i=M),
\end{array}\right.
\left(\begin{array}{c}
1\leq i \leq n-1 \\
0\leq L<M\leq n-1
\end{array}\right)\nonumber
\end{eqnarray}
\begin{eqnarray}
\varphi^{(i)}(z)&=&z^{-1}
\frac{(q^{2n+2}z^2;q^{4n})_\infty}
{(q^{4n}z^2;q^{4n})_\infty}
\frac{(q^{2n}rz;q^{2n})_\infty
(q^{2M}r^{-1}z;q^{2n})_\infty}
{(q^{2n-2M}rz;q^{2n})_\infty
(q^{2M-2L}r^{-1}z;q^{2n})_\infty}
\nonumber\\
&\times&
\frac{(-q^{2n-2i}z^2;q^{4n})_\infty}
{(-q^{2n+2i}z^2;q^{4n})_\infty},~~~
\left(\begin{array}{c}
1\leq i\leq n-1,i\neq L,M\\
0\leq L<M\leq n-1
\end{array}\right)
\label{scalar3}
\end{eqnarray}
and
\begin{eqnarray}
&&\varphi^{(i)}(z)=z^{-1}
\frac
{(q^{2n+2}z^2;q^{4n})_\infty}
{(q^{4n}z^2;q^{4n})_\infty}
\label{scalar4}\\
&\times&
\displaystyle
\frac{(-q^{2(n+L)}z^2;q^{4n})_\infty
(-q^{2(n-i)}z^2;q^{4n})_\infty}
{(-q^{2(n-L)}z^2;q^{4n})_\infty
(-q^{2(n+i)}z^2;q^{4n})_\infty},~~
\left(\begin{array}{c}
1\leq i \leq n-1\\
0\leq L=M \leq n-1
\end{array}\right)
\nonumber
\end{eqnarray}
Let us consider the action of Type-II vertex operators.
Using the bosonic expression of the vacuum $|i\rangle_B$,
we have 
\begin{eqnarray}
\Psi^{*~(i+1,i)}_{n-1}(z)|i\rangle_B
=z^{2(\delta_{i,0}-1)}
\frac{{\varphi}^{(i)}(q^{-1}z^{-1})}
{{\varphi}^{(i)}(q^{-1}z)}
\Psi^{*~(i+1,i)}_{n-1}(z^{-1})|i\rangle_B.
\nonumber
\end{eqnarray}
For $L=0 \leq M \leq n-1$,
A. Doikou and R.I. Nepomechie \cite{DN}
derived the boundary S-matrix by the Bethe Ansatz method.
Their result is expressed by
the $q-$gamma function.
By changing variables
$$
z^{\frac{1}{n}}=(-1)^{\frac{1}{n}}(q^2)^{\sqrt{-1}\lambda/n},
~~~
r=(q^2)^{\xi},
$$
their results coincide to ours.
\subsection{Dual Vacuum}
Let us consider the dual vacuum vector
$~_B\langle i |$.
As the same arguments as the previous subsection,
we construct the bosonic formulas for the dual vacuum
vectors.
We make the ansatz that the dual vacuum has
the following form.
$$
~_B\langle i|=\langle i|e^{G_i},
$$
where 
\begin{eqnarray}
G_i=\sum_{s,t=1}^{n-1}\sum_{k=1}^\infty
\gamma_{s,t}(k)a_s(k)a_{t}(k)+
\sum_{s=1}^{n-1}\sum_{k=1}^\infty
\delta^{(i)}_s(k)a_s(k).
\label{d-bosonic}
\end{eqnarray}
Using the bosonic formulas of the vertex operators,
and comparing the both sides of the equation
(\ref{basic2}),
we have
\begin{eqnarray}
\gamma_{s,t}(k)=\frac{-kq^{-2k}}{2[k]}\times
I_{s,t}(k),
\label{d-result}
\end{eqnarray}
where we have used the element of the matrix 
(\ref{inverse}).
Using the explicit formulas of $\gamma_{s,t}(k)$,
we have the simple formulas of the action
of the basic operators to
the dual vacuum vectors.
\begin{eqnarray*}
~_B\langle i |e^{P^*(z)}&=&
h^{*(i)}(z)_B\langle i |e^{Q^*(1/z)},
\\
~_B\langle i |e^{S_j^-(w)}
&=&g^{*(i)}_j(w)_B\langle i |e^{R_j^-(q^2/w)},~~
(1\leq j \leq n-1),
\end{eqnarray*}
where
\begin{eqnarray*}
h^{*(i)}(z)
&=&
\exp\left(
-\frac{1}{2}\sum_{k=1}^\infty
\frac{[(n-1)k]}{[nk]k}q^kz^{2k}
+\sum_{k=1}^\infty
\frac{[k]}{k}
\delta_{1}^{(i)}(k)
q^{3k/2}z^k\right),\\
\end{eqnarray*}
and
\begin{eqnarray*}
g_j^{*(i)}(w)=
\exp\left(-
\frac{1}{2}\sum_{k=1}^\infty
\frac{[2k]}{[k]k}q^{-k}w^{2k}-
\sum_{k=1}^\infty
\sum_{s=1}^{n-1}\delta_s^{(i)}(k)
\frac{[(a_j|a_s)k]}{k}q^{k/2}w^k\right).
\end{eqnarray*}
The $0$-th component of the 
equation (\ref{basic2}) reduces to
\begin{eqnarray}
\varphi^{(i)}(z)=k_0(z)^{-1}
h^{*(i)}(z),~~(0\leq i \leq n-1).
\end{eqnarray}
Let us cosider the case $~_B\langle 0|$.
As the same arguments as the previous subsection,
the integrand relations reduce to
the following relations.
\begin{eqnarray*}
\frac{g_j^{*(0)}(qw)}
{g_j^{*(0)}(q/w)}=
\left\{\begin{array}{cc}
-w^2,&~j\neq L,M,\\
\displaystyle
-\frac{(1-q^Lrw)}{(1-q^Lr/w)},&~j=L,\\
\displaystyle
-\frac{(1-q^{M-2L}w/r)}
{(1-q^{M-2L}/(rw))},&~j=M,
\end{array}\right.
\left(0\leq L<M\leq n-1\right)
\end{eqnarray*}
Therefore we have
\begin{eqnarray*}
g_j^{*(0)}(qw)=\left\{
\begin{array}{cc}
(1-w^2),&~j\neq L,M,\\
\displaystyle \frac{(1-w^2)}{(1-q^{-L}w/r)},&~j=L,\\
\displaystyle \frac{(1-w^2)}{(1-q^{2L-M}rw)},&~j=M,
\end{array}\right.(0\leq L<M\leq n-1).
\end{eqnarray*}
As the same arguments we have
\begin{eqnarray*}
g_j^{*(0)}(qw)
=\left\{
\begin{array}{cc}
(1-w^2),&~~j\neq L,\\
\displaystyle
\frac{(1-w^2)}{(1+w^2)},&~~j=L,
\end{array}\right.
\left(0\leq L=M \leq n-1\right).
\end{eqnarray*}
Therefore 
the coefficients of the bosonic operators are given by
(\ref{d-result}) and
\begin{eqnarray}
&&\delta_j^{(0)}(k)=-(q^{-k/2}-q^{-3k/2})
\theta_k \sum_{s=1}^{n-1}
\hat{I}_{j,s}(k)\\
&+&
\left\{
\begin{array}{cc}
-q^{(-L-3/2)k}r^{-k}\hat{I}_{j,L}(k)
-q^{(2L-M-3/2)k}r^k\hat{I}_{j,M}(k),&
(0\leq L <M\leq n-1)\\
-2(-1)^{k/2}q^{-3k/2}\theta_k \hat{I}_{j,L}(k),&
(0\leq L=M \leq n-1).
\end{array}\right.\nonumber
\end{eqnarray}
For the other boundary conditions
$~_B\langle i|,~(1\leq i \leq n-1)$,
the integrand relations reduce to the
following relation.
\begin{eqnarray}
\frac{g_j^{*(i)}(qw)}
{g_j^{*(i)}(q/w)}=w^{2\delta_{j,i}}
\frac{g_j^{*(0)}(qw)}
{g_j^{*(0)}(q/w)}.
\end{eqnarray}
Therefore we have
\begin{eqnarray*}
g_j^{*(i)}(qw)=
\left\{\begin{array}{cc}
(1-w^2),&~j\neq L,M,\\
(1-w^2)(1-q^Lrw),&~j=L,\\
\displaystyle
\frac{(1-w^2)}{(1-q^{2L-M}rw)},&j=M,
\end{array}\right.
\left(\begin{array}{c}
1\leq i \leq n-1,~i=L\\
0\leq L<M \leq n-1
\end{array}\right),
\end{eqnarray*}

\begin{eqnarray*}
g_j^{*(i)}(qw)=
\left\{\begin{array}{cc}
(1-w^2),&~j\neq L,M,\\
\displaystyle
\frac{(1-w^2)}{(1-q^{-L}w/r)},&~j=L,\\
(1-w^2)(1-q^{M-2L}w/r),&j=M,
\end{array}\right.
\left(\begin{array}{c}
1\leq i \leq n-1,~i=M\\
0\leq L<M \leq n-1
\end{array}\right),
\end{eqnarray*}

\begin{eqnarray*}
g_j^{*(i)}(qw)=
\left\{\begin{array}{cc}
(1-w^2),&~j\neq L,M,i,\\
(1-w^2)(1+w^2),&~j=i,\\
\displaystyle
\frac{(1-w^2)}{(1-q^{-L}w/r)},&~j=L,\\
\displaystyle
\frac{(1-w^2)}{(1-q^{2L-M}rw)},&j=M,
\end{array}\right.
\left(\begin{array}{c}
1\leq i \leq n-1,~i\neq L,M\\
0\leq L<M \leq n-1
\end{array}\right),
\end{eqnarray*}
and
\begin{eqnarray*}
g_j^{*(i)}(qw)=
\left\{\begin{array}{cc}
(1-w^2),&~j\neq L,i,\\
(1-w^2)(1+w^2),&~j=i,\\
\displaystyle
\frac{(1-w^2)}{(1+w^2)},&~j=L,
\end{array}\right.
\left(\begin{array}{c}
1\leq i \leq n-1,~i\neq L\\
0\leq L=M \leq n-1
\end{array}\right),
\end{eqnarray*}

\begin{eqnarray*}
g_j^{*(i)}(qw)=(1-w^2),~\left(
\begin{array}{c}
0\leq i \leq n-1\\
0\leq L=M=i \leq n-1
\end{array}\right)
\end{eqnarray*}
Therefore the coefficients of the bosonic operators 
are given by
(\ref{d-result}) and 

\begin{eqnarray}
&&\delta_j^{(i)}(k)=-(q^{-k/2}-q^{-3k/2})
\theta_k \sum_{s=1}^{n-1}
\hat{I}_{j,s}(k)\label{d-resulti}\\
&+&\left\{\begin{array}{cc}
q^{(L-3/2)k}r^k \hat{I}_{j,L}(k)
-q^{(2L-M-3/2)k}r^k \hat{I}_{j,M}(k),&i=L\\
-q^{(-L-3/2)k}r^{-k} \hat{I}_{j,L}(k)
+q^{(M-2L-3/2)k}r^{-k} \hat{I}_{j,M}(k),&i=M\\
-q^{(-L-3/2)k}r^{-k} \hat{I}_{j,L}(k)
-q^{(2L-M-3/2)k}r^k \hat{I}_{j,M}(k)&\\
+2(-1)^{k/2}q^{-3k/2}\theta_k \hat{I}_{j,i}(k),&
i\neq L,M
\end{array}\right.
\left(\begin{array}{c}
1\leq i \leq n-1\\
0\leq L<M \leq n-1
\end{array}\right)
\nonumber
\end{eqnarray}
and
\begin{eqnarray}
&&\delta_j^{(i)}(k)=-(q^{-k/2}-q^{-3k/2})
\theta_k\sum_{s=1}^{n-1}\hat{I}_{j,s}(k)\nonumber\\
&+&
2(-1)^{k/2}q^{-3k/2}\theta_k
(\hat{I}_{j,i}(k)-\hat{I}_{j,L}(k)),~~
\left(\begin{array}{c}
1\leq i \leq n-1 \\
0\leq L=M \leq n-1
\end{array}\right).
\end{eqnarray}
The scalar factors of the refrection matrix
$\varphi^{(i)}(z)$ are given by
\begin{eqnarray}
\varphi^{(i)}(z)=\frac{1}{k_0(z)}
\frac{(q^{2n+2}z^2;q^{4n})_\infty}
{(q^{4n}z^2;q^{4n})_\infty},~~~(0\leq L=M=i \leq n-1)
\label{scalar5},
\end{eqnarray}
\begin{eqnarray}
&&\varphi^{(0)}(z)=\frac{1}{k_0(z)}
\frac{(q^{2n+2}z^2;q^{4n})_\infty}{(q^{4n}z^2;q^{4n})_\infty}
\label{scalar6}\\
&\times&
\left\{\begin{array}{cc}
\displaystyle
\frac{(rq^{2L}z;q^{2n})_\infty 
((r^{-1}z;q^{2n})_\infty)^{1-\delta_{L,0}}}
{(rq^{2n-2M+2L}z;q^{2n})_\infty 
((r^{-1}q^{2n-2L}z;q^{2n})_\infty)^{1-\delta_{L,0}}},&
~~(0\leq L < M \leq n-1),\\
\displaystyle
\frac{((-q^{2L}z^2;q^{4n})_\infty)^{1-\delta_{L,0}}}
{((-q^{4n-2L}z^2;q^{4n})_\infty)^{1-\delta_{l,0}}},&
~~(0\leq L=M \leq n-1).\end{array}\right.
\nonumber
\end{eqnarray}

\begin{center}
\begin{eqnarray}
&&\varphi^{(i)}(z)=\frac{1}{k_0(z)}
\frac{(q^{2n+2}z^2;q^{4n})_\infty}{(q^{4n}z^2;q^{4n})_\infty}
\label{scalar7}\\
&\times&
\left\{\begin{array}{c}
\displaystyle
\frac{(rq^{2n}z;q^{2n})_\infty}
{(rq^{2n-2M+2L}z;q^{2n})_\infty},~(1\leq L=i < M \leq n-1),\\
\displaystyle
\frac{(r^{-1}q^{2n-2L}z;q^{2n})_\infty 
((r^{-1}z;q^{2n})_\infty)^{1-\delta_{L,0}}}
{(r^{-1}q^{2M-2L}z;q^{2n})_\infty 
((r^{-1}q^{2n-2L}z;q^{2n})_\infty)^{1-\delta_{L,0}}},
(0\leq L < M=i \leq n-1),\\
\displaystyle
\frac{(-q^{4n-2i}z^2;q^{4n})_\infty (rq^{2L}z;q^{2n})_\infty
((r^{-1}z;q^{2n})_\infty)^{1-\delta_{L,0}}}
{(-q^{2i}z^2;q^{4n})_\infty (rq^{2n-2M+2L}z;q^{2n})_\infty 
((r^{-1}q^{2n-2L}z;q^{2n})_\infty)^{1-\delta_{L,0}}},\\
~~~~~~~~~(0\leq L < M \leq n-1 , i \neq L,M),\\
\displaystyle
\frac{(-q^{4n-2i}z^2;q^{4n})_\infty 
((-q^{2L}z^2;q^{4n})_\infty)^{1-\delta_{L,0}}}
{(-q^{2i}z^2;q^{4n})_\infty 
((-q^{4n-2L}z^2;q^{4n})_\infty)^{1-\delta_{l,0}}},
(0\leq L=M \leq n-1 , i \neq L).\end{array}\right.
\nonumber
\end{eqnarray}
\end{center}

\section{Correlation functions}

In the previous section we have constructed
the bosonic formulas of the vacuum and the dual vacuum.
In this section we consider an application of
these bosonic formulas.
We have constructed both vacuum and dual vacuum
for the same transfer matrix,
in the following cases.
$$
(1)~\Lambda_i,~~(0\leq i\leq n-1),~0\leq L=M=i \leq n-1,
$$
$$
(2)~\Lambda_i,~(0\leq i \leq n-2),~0\leq L=i<M\leq n-1,
$$
$$
(3)~\Lambda_i,~(1\leq i \leq n-1),~0\leq L<M=i\leq n-1.
$$
Therefore we can derive the vacuum expectation value
for the above cases.

Let $L$ be a linear opertor on the $m$-fold tensor
product of the $n$-dimensional vector space
$V \otimes \cdots \otimes V$.
The corresponding local operator ${\mathcal{L}}$
acting on our space of states $V(\Lambda_i)$
can be defined in terms of the Type-I vertex operators,
in exactly the same way
as in the bulk theory \cite{JM}.
Explicitly, if $E_{j,k}(m)$ is the spin operator at
the $m$-site
$$
{{E_{j,k}(m)}=E_{j,k}\otimes id \otimes \cdots \otimes id},
$$
the corresponding local operator
${\mathcal{E}_{j,k}^{(i)}(m)}$
is given by
\begin{eqnarray}
&&\mathcal{E}_{j,k}^{(i)}(m)
=g_n^m
\sum_{j_1 \cdots j_{m-1}=0}^{n-1}
\Phi_{j_1}^{*(i,i-1)}(1)\cdots
\Phi_{j_{m-1}}^{*(i-m+2,i-m+1)}(1)\nonumber\\
&\times&
\Phi_{j}^{*(i-m+1,i-m)}(1)
\Phi_{k}^{(i-m,i-m+1)}(1)
\Phi_{j_{m-1}}^{(i-m+1,i-m+2)}(1)\cdots
\Phi_{j_{1}}^{(i-1,i)}(1),\nonumber
\end{eqnarray}
where we have used
$$
g_n=\frac{(q^2;q^{2n})_\infty}{(q^{2n};q^{2n})_\infty}.
$$
Therefore the boundary magnetization is given by
\begin{eqnarray}
\sum_{j=0}^{n-1}\omega^j P^{(i)}_j(1),
\label{magnetization}
\end{eqnarray} 
where $\omega$ is an n-th primitive root of $1$, and
we have used the one-point function $P_j^{(i)}(z)$
with a spectral parameter $z$, defined by
\begin{eqnarray}
P^{(i)}_j(z)=g_n
\frac{~_B\langle i |
\Phi_j^{*~(i,i-1)}(z)\Phi_j^{(i-1,i)}(z)|i\rangle_B}
{~_B\langle i|i\rangle_B}.
\label{expectation}
\end{eqnarray}
In order to evaluate the expectation value
(\ref{expectation}),
we invoke the bosonization formulas of various quantities.
By normal-ordering the product of 
vertex operators, for $j=0,n-1$, we have
\begin{eqnarray}
&&P_j^{(i)}(z)=q^i
(1-q^2)^n
\oint\frac{dw_1}{2\pi i w_1}\cdots
\oint\frac{dw_{n-1}}{2\pi i w_{n-1}}\frac{1}{w_i}
\nonumber \\
&\times&
\prod_{l=1}^{n-2}
\frac{1}{(1-qw_l/w_{l+1})(1-qw_{l+1}/w_l)}
I(z,w_1/q,\cdots,w_j/q,q^{n+1}w_{j+1}
,\cdots,q^{n+1}w_{n-1})
\nonumber
\\
&\times&
\left\{
\begin{array}{cc}
\displaystyle
\frac{w_1}{z}\frac{1}{(1-q^{n+1}w_1/z)
(1-qz/w_{n-1})(1-qw_{n-1}/z)},&~~(j=0),\\
\displaystyle
w_{n-1}\frac{1}{(1-qz/w_1)(1-qw_1/z)
(1-q^{n+3}z/w_{n-1})},&~~(j=n-1).
\end{array}\right.\label{integ1}
\end{eqnarray}
where the contours of integrals are taken as
$|qw_{l}/w_{l+1}|, 
|qw_{l+1}/w_l|<1,~(1\leq l \leq n-2)$
and, for $j=0$ case we add the conditions
$|q^{n+1}w_1/z|,
|qz/w_{n-1}|,|qw_{n-1}/z|<1$,
for $j=n-1$ case,
we add the conditions
$|qz/w_{1}|, |qw_1/z|, |q^{n+3}z/w_{n-1}|<1$.\\
For $1\leq j \leq n-2$, we have
\begin{eqnarray}
&&P_j^{(i)}(z)=q^i(1-q^2)^n
\oint\frac{dw_1}{2\pi i w_1}\cdots
\oint\frac{dw_{n-1}}{2\pi i w_{n-1}}\frac{1}{w_i}
\label{integ2}\\
&\times&
\prod_{l=1
\atop{l\neq j}}^{n-2}
\frac{1}{(1-qw_l/w_{l+1})(1-qw_{l+1}/w_l)}
I(z,w_1/q,\cdots,w_j/q,q^{n+1}w_{j+1}
,\cdots,q^{n+1}w_{n-1})\nonumber
\\
&\times&
\frac{w_jw_{j+1}}{z}
\frac{1}{(1-qz/w_1)(1-qw_1/z)
(1-q^{n+3}w_{j+1}/w_{j})(1-qz/w_{n-1})(1-qw_{n-1}/z)},
\nonumber
\end{eqnarray}
where the contour of integrals
is taken as $|qw_l/w_{l+1}|,
|qw_{l+1}/w_l|<1,~~(1\leq l\neq j\leq n-2)$ 
and
$|qz/w_1|,|qw_1/z|$,
$|q^{n+3}w_{j+1}/w_{j}|,
|qz/w_{n-1}|, |qw_{n-1}/z|<1$.\\
Here we have set
\begin{eqnarray}
&&I(z,w_1,\cdots,w_{n-1})
\times
~_B\langle i |i\rangle_B\nonumber
\\
=&&
\displaystyle~_B\langle i |
\exp\left([2k]
\sum_{k=1}^\infty
\sum_{p=1}^{n-1}
\sum_{l=1}^{n-1}I_{p,l}(k)x_l(k)a_p(-k)
\right)\nonumber
\\
&&~~~~\times
\exp\left([2k]
\sum_{k=1}^\infty \sum_{p=1}^{n-1}
\sum_{l=1}^{n-1}I_{p,l}(k)y_l(k)a_p(k)\right)
|i\rangle_B,
\end{eqnarray}
where 
\begin{eqnarray}
x_j(k)=
\left\{
\begin{array}{cc}
\displaystyle
\frac{q^{3k/2}}{[2k]}z^k-\frac{q^{k/2}}{[k]}
w_1^k+\frac{q^{k/2}}{[2k]}w_2^k,&~~(j=1),\\
\displaystyle
\frac{q^{k/2}}{[2k]}w_{j-1}^k-\frac{q^{k/2}}{[k]}
w_j^k+\frac{q^{k/2}}{[2k]}w_{j+1}^k,&~~(2\leq j \leq n-2),\\
\displaystyle
\frac{q^{(2n+3)k/2}}{[2k]}z^k-\frac{q^{k/2}}{[k]}
w_{n-2}^k+\frac{q^{k/2}}{[2k]}w_{n-1}^k,&~~(j=n-1),
\end{array}
\right.
\end{eqnarray}
and
\begin{eqnarray}
y_j(k)=
\left\{
\begin{array}{cc}
\displaystyle
-\frac{q^{-k/2}}{[2k]}z^{-k}+\frac{q^{k/2}}{[k]}
w_1^{-k}-\frac{q^{k/2}}{[2k]}w_2^{-k},&~~(j=1),\\
\displaystyle
-\frac{q^{k/2}}{[2k]}w_{j-1}^{-k}+\frac{q^{k/2}}{[k]}
w_j^{-k}-\frac{q^{k/2}}{[2k]}w_{j+1}^{-k},&~~(2\leq j \leq n-2),\\
\displaystyle
-\frac{q^{-(2n+1)k/2}}{[2k]}z^{-k}
+\frac{q^{k/2}}{[k]}
w_{n-2}^{-k}-\frac{q^{k/2}}{[2k]}w_{n-1}^{-k},&~~(j=n-1),
\end{array}
\right.
\end{eqnarray}
Here $I_{p,l}(k)$ is defined in (\ref{inverse}).
To calculate the vacuum expectation values
(\ref{Integrand}), 
we use the coherent states.
Let us define the coherent state by
\begin{eqnarray*}
|\xi_1\cdots \xi_{n-1}\rangle_i=
\exp\left(
\sum_{p=1}^{n-1}\sum_{k=1}^\infty
\frac{k}{[k][2k]}\xi_p(k)a_p(-k)\right)|i\rangle,
\end{eqnarray*}
and
\begin{eqnarray*}
~_i\langle \bar{\xi}_1 \cdots \bar{\xi}_{n-1}|
=\langle i|\exp\left(
\sum_{p=1}^{n-1}\sum_{k=1}^\infty
\frac{k}{[k][2k]} \bar{\xi}_p(k)a_p(k)\right).
\end{eqnarray*}
The coherent states enjoy
\begin{eqnarray*}
a_p(k)|\xi_1 \cdots \xi_{n-1}\rangle_i=
\sum_{j=1}^{n-1}\frac{[(a_p|a_j)k]}{[2k]}\xi_j(k)|
\xi_1 \cdots \xi_{n-1}\rangle_i,\\
~_i\langle \bar{\xi}_1 \cdots \bar{\xi}_{n-1}|a_p(-k)
=_i\langle \bar{\xi}_1 \cdots \bar{\xi}_{n-1}|
\sum_{j=1}^{n-1}\frac{[(a_p|a_j)k]}{[2k]}\bar{\xi}_j(k),
\end{eqnarray*}
and
\begin{eqnarray*}
id&=&
\int_{-\infty}^\infty
\left(\frac{-1}{2\pi i}\right)^{n-1}
\prod_{i=1}^{n-1}\prod_{k>0}
\frac{k[(i+1)k]d\xi_i(k)d\bar{\xi}_i(k)}{[2k]^2[k]}\\
&\times&
\exp\left(-\sum_{i,j=1}^{n-1}\sum_{k=1}^\infty
\frac{[(a_i|a_j)k]k}{[k][2k]^2}
\xi_i(k)\bar{\xi}_j(k)\right)|\xi_1 \cdots \xi_{n-1}\rangle_i
~_i\langle \bar{\xi}_1 \cdots \bar{\xi}_{n-1}|,
\end{eqnarray*}
where the integration is taken over the entire complex plane
with the measure 
$d\xi d\bar{\xi}=-2 i dxdy$ for $\xi=x+iy$.
Using this completness relation, we have
the following.
\begin{eqnarray*}
&&I(z,w_1,\cdots,w_{n-1})
\\
&=&
\prod_{k=1}^\infty
\exp\left(
\frac{1}{q^{2nk}-1}\frac{[2k]^2}{[k][nk]k}
\right.
\\
&\times&
\left\{
\sum_{l=1}^{n-1}
[lk][(n-l)k]
\left(-q^{2nk}x_l(k)y_l(k)+
\frac{q^{-2k}}{2}x_l(k)^2+
\frac{q^{2(n+1)k}}{2}y_l(k)^2\right)\right.\\
&+&
\sum_{1\leq l_1 <l_2 \leq n-1}
[l_1k][(n-l_2)k]\left(
-q^{2nk} x_{l_1}y_{l_2}
-q^{2nk} x_{l_2}y_{l_1}
+q^{-2k}x_{l_1}x_{l_2}
+q^{2(n+1)k}y_{l_1}y_{l_2}\right)\nonumber
\\
&+&
\sum_{l=1}^{n-1}[lk][(n-l)k]
\left(\bar{\beta}^{(i)}_l(k)
(q^{-2k}x_{l}(k)-y_{l}(k))
+\bar{\delta}_l^{(i)}(k)
(q^{2(n+1)k}y_l(k)-x_l(k))\right)\\
&+&
\sum_{1\leq l_1 <l_2 \leq n-1}
[l_1k][(n-l_2)k]\left(
\bar{\beta}^{(i)}_{l_1}(k)(q^{-2k}x_{l_2}(k)-y_{l_2}(k))
+\bar{\beta}^{(i)}_{l_2}(k)(q^{-2k}x_{l_1}(k)-y_{l_1}(k))
\right.\\
&+&\left.\left.\left.
\bar{\delta}_{l_1}^{(i)}(k)
(q^{2(n+1)k}y_{l_2}(k)-x_{l_2}(k))+
\bar{\delta}_{l_2}^{(i)}(k)
(q^{2(n+1)k}y_{l_1}(k)-x_{l_1}(k))\right)\right\}\right).
\end{eqnarray*}
and
\begin{eqnarray*}
&&~_B\langle i | i\rangle_B\\
&=&\prod_{k=1}^\infty
\left(\frac{1}{\sqrt{1-q^{2nk}}}\right)^{n-1}
\prod_{k=1}^\infty
\exp\left(\frac{1}{q^{2nk}-1}\frac{[2k]^2}{[k][nk]k}
\right.
\\
&&\left\{
\sum_{l=1}^{n-1}[lk][(n-l)k]\left(
-\bar{\beta}_{l}^{(i)}(k)\bar{\delta}_{l}^{(i)}(k)
+\frac{q^{-2k}}{2}\bar{\beta}_l^{(i)}(k)^2
+\frac{q^{2(n+1)k}}{2}\bar{\delta}_{l}^{(i)}(k)^2\right)\right.\\
&+&\left.\left.
\sum_{1\leq l_1<l_2 \leq n-1}
[l_1k][(n-l_2)k]\left(
-\bar{\beta}_{l_1}^{(i)}\bar{\delta}_{l_2}^{(i)}
-\bar{\delta}_{l_1}^{(i)}\bar{\beta}_{l_2}^{(i)}
+q^{-2k}\bar{\beta}_{l_1}^{(i)}\bar{\beta}_{l_2}^{(i)}
+q^{2(n+1)k}\bar{\delta}_{l_1}^{(i)}
\bar{\delta}_{l_2}^{(i)}\right)\right\}\right).
\end{eqnarray*}
Here we have used
$$
\bar{\beta}_l^{(i)}(k)=\sum_{s=1}^{n-1}
\frac{[(a_l|a_s)k]}{[2k]}\beta_s^{(i)}(k),~~
\bar{\delta}_l^{(i)}(k)=\sum_{s=1}^{n-1}
\frac{[(a_l|a_s)k]}{[2k]}\delta_s^{(i)}(k).
$$
The sum in the right-hand sides are evaluated as follows.

~\\
{\it The Norm of the vacuum vectors :}\\
~\\
(1)~$\Lambda_i,~(0\leq i \leq n-1),~
0\leq L=M=i \leq n-1$ case.
\begin{eqnarray}
&&~_B\langle i | i \rangle_B
\label{norm1}
\\
&=&
\frac{1}{\sqrt{(q^{4n};q^{4n})_\infty}}\nonumber
\prod_{j=1}^{n-1}\left\{
\frac{\sqrt{(q^{4n+2-2j};q^{4n})_\infty 
(q^{4n-2-2j};q^{4n})_\infty}}
{(q^{4n-2j};q^{4n})_\infty}\right\}^{j(n-j)}.
\end{eqnarray}
(2)~$\Lambda_i,~(0\leq i \leq n-2),~
0\leq L=i< M\leq n-1$ case.
\begin{eqnarray}
&&~_B\langle i | i \rangle_B
\label{norm2}
\\
&=&
\frac{1}{\sqrt{(q^{4n};q^{4n})_\infty}}\nonumber
\prod_{j=1}^{n-1}\left\{
\frac{\sqrt{(q^{4n+2-2j};q^{4n})_\infty 
(q^{4n-2-2j};q^{4n})_\infty}}
{(q^{4n-2j};q^{4n})_\infty}\right\}^{j(n-j)}\nonumber\\
&\times&
\displaystyle
\prod_{s=1}^{M-L}
\frac{(q^{4n-2M+2L-2s}r^2;q^{4n})_\infty}
{(q^{2n-2s}r^2;q^{4n})_\infty},~~~(0\leq L=i<M\leq n-1),
\nonumber
\end{eqnarray}
(3)~$\Lambda_i,~(1\leq i \leq n-1),~
0\leq L< M=i\leq n-1$ case.
\begin{eqnarray}
&&~_B\langle i | i \rangle_B
\label{norm3}
\\
&=&
\frac{1}{\sqrt{(q^{4n};q^{4n})_\infty}}\nonumber
\prod_{j=1}^{n-1}\left\{
\frac{\sqrt{(q^{4n+2-2j};q^{4n})_\infty 
(q^{4n-2-2j};q^{4n})_\infty}}
{(q^{4n-2j};q^{4n})_\infty}\right\}^{j(n-j)}
\nonumber\\
&\times&
\displaystyle
\prod_{s=1}^{M-L}
\frac{(q^{2n+2M-2L-2s}r^{-2};q^{4n})_\infty}
{(q^{4M-4L-2s}r^{-2};q^{4n})_\infty},~~~
(0\leq L<M=i\leq n-1).
\nonumber
\end{eqnarray}
~\\
{\it Integrand of Correlation Functions :}\\
~\\
(1)~$\Lambda_i,~(0\leq i \leq n-1),~
0\leq L=M=i \leq n-1$ case.
\begin{eqnarray}
I(z,w_1,\cdots, w_{n-1})
=J(z,w_1,\cdots,w_{n-1}),~(0\leq L=M=i \leq n-1).
\label{Integrand1}
\end{eqnarray}
(2)~$\Lambda_i,~(0\leq i \leq n-2),~
0\leq L=i<M\leq n-1$ case.
\begin{eqnarray}
&&I(z,w_1,\cdots, w_{n-1})
\label{Integrand2}\\
&=&J(z,w_1,\cdots,w_{n-1})\times
\displaystyle 
\frac{(1-rz)}{
(1-rq^{-M-1}w_M)},~
\left(\begin{array}{c}
i=L=0<M\leq n-1
\end{array}\right),\nonumber
\end{eqnarray}
and
\begin{eqnarray}
&&I(z,w_1,\cdots,w_{n-1})=J(z,w_1,\cdots,w_{n-1})
\label{Integrand3}\\
&\times&
\frac
{(q^{L-1}rw_L)_\infty
(q^{2n+L+1}r/w_L)_\infty
(q^{2n-M-1}rw_M)_\infty
(q^{2n-M+1}r/w_M)_\infty}
{(q^{2n-2M+L-1}rw_L)_\infty
(q^{2n-2M+L+1}r/w_L)_\infty
(q^{2L-M-1}rw_M)_\infty
(q^{2n+2L-M+1}r/w_M)_\infty
}\nonumber\\
&\times& 
\frac{(q^{2n+2L-2M}rz)_\infty 
(q^{2n+2L-2M}rz^{-1})_\infty}
{(q^{2n}rz)_\infty 
(q^{2n}rz^{-1})_\infty},
~~\left(1\leq L=i <M\leq n-1\right).\nonumber
\end{eqnarray}
(3)~$\Lambda_i~(1\leq i\leq n-1),~0\leq L<M=i \leq n-1.$
\begin{eqnarray}
&&I(z,w_1,\cdots,w_{n-1})
\label{Integrand4}\\
&=&J(z,w_1,\cdots,w_{n-1})
\times \frac{(1-1/(rz))}{(1-q^{M+1}/(rw_M))},~
(0=L<M=i\leq n-1),\nonumber
\end{eqnarray}
and
\begin{eqnarray}
&&I(z,w_1,\cdots,w_{n-1})
=J(z,w_1,\cdots,w_{n-1})\label{Integrand5}\\
&\times&
\frac{(q^{2M-L-1}r^{-1}w_L)_\infty 
(q^{2M-L+1}r^{-1}w_L^{-1})_\infty
(q^{M-2L-1}r^{-1}w_M)_\infty
(q^{2n-2L+M+1}r^{-1}w_M^{-1})_\infty}
{(q^{-L-1}r^{-1}w_L)_\infty 
(q^{2n-L+1}r^{-1}w_L^{-1})_\infty
(q^{M-1}r^{-1}w_M)_\infty
(q^{M+1}r^{-1}w_M^{-1})_\infty}\nonumber\\
&\times&
\frac{(r^{-1}z)_\infty (r^{-1}z^{-1})_\infty}
{(q^{2M-2L}r^{-1}z)_\infty 
(q^{2M-2L}r^{-1}z^{-1})_\infty},~~(1\leq L<M=i\leq n-1).
\nonumber
\end{eqnarray}
Here we have set the function 
$J(z,w_1,\cdots,w_{n-1})$ by
\begin{eqnarray}
&&J(z,w_1,\cdots,w_{n-1})
=(q^{2n})_\infty^n
(q^{2n+2})_\infty^n
\frac{(q^nz;q^n)_\infty(q^nz^{-1};q^n)_\infty}
{(qz;q^n)_\infty(qz^{-1};q^n)_\infty}
\nonumber\\
&\times&
\sqrt{(q^{2n}z^2)_\infty
(q^{2n}z^{-2})_\infty
(q^2z^2)_\infty
(q^{2}z^{-2})_\infty}
\label{Integrand}\\
&\times&
\prod_{j=1}^{n-1}\left\{
\frac{(q^{-1}w_j;q^n)_\infty(q^{n+1}w_j^{-1};q^n)_\infty}
{(w_j;q^n)_\infty(q^{n+2}w_j^{-1};q^n)_\infty}
\sqrt{(w_j^2)_\infty
(q^{-2}w_j^2)_\infty
(q^{2n}w_j^{-2})_\infty
(q^{2n+2}w_j^{-2})_\infty}\right\}
\nonumber\\
&\times&
\prod_{j=1}^{n-2}
\left\{
(q^{-1}w_jw_{j+1})_\infty
(q^{2n+1}w_jw_{j+1}^{-1})_\infty
(q^{2n+1}w_j^{-1}w_{j+1})_\infty
(q^{2n+3}w_j^{-1}w_{j+1}^{-1})_\infty
\right\}^{-1}\nonumber\\
&\times&
\left\{(zw_1)_\infty(q^{2n+2}zw_1^{-1})_\infty
(q^{2n}z^{-1}w_1)_\infty(q^{2n+2}z^{-1}w_1^{-1})_\infty
\right\}^{-1}\nonumber \\
&\times&
\left\{(q^nzw_{n-1})_\infty
(q^{3n+2}zw_{n-1}^{-1})_\infty
(q^{n}z^{-1}w_{n-1})_\infty(q^{n+2}z^{-1}w_{n-1}^{-1})_\infty
\right\}^{-1}.\nonumber 
\end{eqnarray} 
Here we have used the abberiviation.
$$(z)_\infty=(z;q^{2n})_\infty.$$

{\it We summarize the main result of this section.
For the asymptotic boundary conditions
$V(\Lambda_i)~,(i=0,\cdots,n-1)$,
let us consider the following boundary conditions,}
$$
(1)~
0\leq L=M=i \leq n-1,
~~
(2)~
0\leq L=i<M \leq n-1,
~~
(3)~
0\leq L<M=i \leq n-1.
$$
{\it Then the boundary magnetization is given by}
$$
\sum_{j=0}^{n-1}\omega^j P_j^{(i)}(1),
$$
{\it where the correlation function}
$P_j^{(i)}(z)$ {\it is given in} (\ref{integ1}) and
(\ref{integ2}),
{\it the integrand function of the correlation
function} $P_j^{(i)}(z)$ {\it is given in}
(\ref{Integrand1}), (\ref{Integrand2}),
(\ref{Integrand3}),(\ref{Integrand4}),
(\ref{Integrand5}), and (\ref{Integrand}).
{\it and} $\omega$ {\it is an n-th primitive root of} $1$.

~\\
{\sl Acknowledgements}~~~~~~~
We want to thank to Professor M. Jimbo,
Professor T. Miwa and Professor R. Nepomechie for
their interests to this work.
We want to thank to S. Yamasita for useful discussion,
to improve our manuscript.
This work is partly
supported by the Grant from Research Institute
of Science and Technology, Nihon University,
and the Grant from the Ministry of Education,
Science, Sports, and Culture, Japan
(11740099).

\section{Appendix}
For reader's convenience, we summarize the bosonizations
of the vertex operators \cite{Koy}.
Let ${\mathbb C}[\bar{P}]$
be the ${\mathbb C}$-algebra generated by symbols
$\{e^{\alpha_2},\cdots,e^{\alpha_{n-1}},
e^{\bar{\Lambda}_{n-1}}\}$
which satisfy the following defining relations :
\begin{eqnarray*}
e^{\alpha_i}e^{\alpha_j}=(-1)^{(\alpha_i|\alpha_j)}
e^{\alpha_j}e^{\alpha_i},~~(2\leq i,j \leq n-1),\\
e^{\alpha_i}e^{\bar{\Lambda}_{n-1}}=
(-1)\delta_{i,n-1}e^{\bar{\Lambda}_{n-1}}
e^{\alpha_i},~~(2\leq i \leq n-1).
\end{eqnarray*}
For $\alpha=m_2 \alpha_2+\cdots +m_{n-1}\alpha_{n-1}
+m_n \bar{\Lambda}_{n-1}$, we denote
$e^{m_2 \alpha_2}\cdots e^{m_{n-1}\alpha_{n-1}}
e^{m_n \bar{\Lambda}_{n-1}}$
by $e^{\alpha}$.
Let ${\mathbb C}[\bar{Q}]$ be the ${\mathbb C}$-algebra
generated by the symbols
$\{e^{\alpha_1}, \cdots, e^{\alpha_{n-1}}\}$
which satisfy the following defining relations :
\begin{eqnarray*}
e^{\alpha_i}e^{\alpha_j}=
(-1)^{(\alpha_i|\alpha_j)}e^{\alpha_j}
e^{\alpha_i},
~~(1\leq i,j \leq n-1).
\end{eqnarray*}
Let the boson be the ${\mathbb C}$-algebra
generated by the symbols
$a_s(k),~(s\in \{0,1,\cdots, n-1 \}, k \in {\mathbb Z})$ 
which satisfy the following defining relations :
$$
[a_s(k),a_t(l)]=\delta_{s,t}\frac{[(a_s|a_t)k][k]}{k}.
$$
The highset weight module $V(\Lambda_i)$ is realized as
$$
V(\Lambda_i)={\mathbb C}[a_s(-k),
~(s\in \{0,1,\cdots,n-1\},k \in {\mathbb Z}\geq 0)]
\otimes {\mathbb C}[\bar{Q}]e^{\bar{\Lambda}_i}.
$$
Here the actions of the operators
$a_s(k), \partial_{\alpha}, e^{\alpha}$
on $V(\Lambda_i)$
are defined as follows :

\begin{eqnarray*}
a_s(k) f \otimes e^{\beta}=
\left\{
\begin{array}{cc}
a_s(k) f \otimes e^{\beta}, &~(k<0),\\
\left[a_s(k),f\right] \otimes e^{\beta},&~(k>0),
\end{array}\right.
\end{eqnarray*}
\begin{eqnarray*}
\partial_{\alpha} f \otimes e^{\beta}&=&(\alpha|\beta)
f \otimes e^{\beta}.\\
e^{\alpha} f \otimes e^{\beta}&=&f\otimes e^{\alpha}
e^{\beta}.
\end{eqnarray*}
The bosonizations of the vertex operators are given by 
\begin{eqnarray*}
\Phi_{n-1}^{(i,i+1)} (z) &=& e^{P(z)}e^{Q(z)}e^{\bar{\Lambda}_{n-1}}
(q^{n+1}z)^{\partial_{\bar{\Lambda}_{n-1}}+\frac{n-i-1}{n}}
(-1)^{(\partial_{\bar{\Lambda}_1}-\frac{n-i-1}{n})(n-1)+\frac{1}{2}(n-i)(n-i-1)},\\
\Phi_{j-1}^{(i,i+1)} (z) &=& [ \Phi_j^{(i,i+1)} (z) , f_j ]_q
= \oint \frac{dw_j}{2\pi i} [ \Phi_j^{(i,i+1)} (z) ,
e^{R_j^{-}(w_j)}e^{S_j^{-}(w_j)}e^{-\alpha_j}w_j^{-\partial_{\alpha_j}}
]_q,
\\
\Phi_0^{* (i+1,i)} (z) &=& e^{P^*(z)}e^{Q^*(z)}e^{\bar{\Lambda}_1}
((-1)^{n-1}qz)^{\partial_{\bar{\Lambda}_1}+\frac{i}{n}}q^i
(-1)^{in+\frac{1}{2}i(i+1)},\\
\Phi_j^{* (i+1,i)} (z) &=& [ f_j , \Phi_{j-1}
^{* (i+1,i)}(z) ]_{q^{-1}}
= \oint \frac{dw_j}{2\pi i} [
e^{R_j^{-}(w_j)}e^{S_j^{-}(w_j)}e^{-\alpha_j}w_j^{-\partial_{\alpha_j}}
, \Phi_{j-1}^{* (i+1,i)}(z) ]_{q^{-1}},
\\
\Psi_0^{(i,i+1)} (z) &=& e^{-P^*(q^{-1}z)}e^{-Q^*(qz)}e^{-\bar{\Lambda}_1}
((-1)^{n+1}qz)^{-\partial_{\bar{\Lambda}_1}+\frac{n-i-1}{n}}
q^{-i}(-1)^{in+\frac{1}{2}i(i+1)},\\
\Psi_j^{(i,i+1)} (z) &=& [ \Psi_{j-1}^{(i,i+1)} (z) , e_j ]_q
= \oint \frac{dw_j}{2\pi i} [ \Psi_{j-1}^{(i,i+1)} (z) ,
e^{-R_j^{-}(q^{-1}w_j)}e^{-S_j^{-}(qw_j)}e^{\alpha_j}w_j^{\partial_{\alpha_j}}
]_q,
\\
\Psi_{n-1}^{* (i+1,i)} (z) &=& e^{-P(q^{-1}z)}e^{-Q(qz)}e^{-\bar{\Lambda}_{n-1}}
(q^{n+1}z)^{-\partial_{\bar{\Lambda}_{n-1}}+\frac{i}{n}}
(-1)^{(\partial_{\bar{\Lambda}_1}-\frac{n-i}{n})(n-1)+\frac{1}{2}(n-i)(n-i-1)},\\
\Psi_{j-1}^{* (i+1,i)} (z) &=& [ e_j , \Psi_j^{* (i+1,i)}(z) ]_{q^{-1}}
= \oint \frac{dw_j}{2\pi i} [
e^{-R_j^{-}(q^{-1}w_j)}e^{-S_j^{-}(qw_j)}e^{\alpha_j}w_j^{\partial_{\alpha_j}}
, \Psi_j^{* (i+1,i)}(z) ]_{q^{-1}},
\\
&&(i=0,\cdots,n-1, \hspace{1cm} j=1,\cdots,n-1),
\end{eqnarray*}
where we have used 
\begin{eqnarray*}
P(z) = \sum_{k=1}^{\infty} a_{n-1}^*(-k)q^{\frac{2n+3}{2}k}z^k, &&
Q(z) = \sum_{k=1}^{\infty} a_{n-1}^*(k)q^{-\frac{2n+1}{2}k}z^{-k},\\
P^*(z) = \sum_{k=1}^{\infty} a_1^*(-k)q^{\frac{3}{2}k}z^k, &&
Q^*(z) = \sum_{k=1}^{\infty} a_1^*(k)q^{-\frac{1}{2}k}z^{-k},\\
R_j^{-}(w) = -\sum_{k=1}^{\infty}
\frac{a_j(-k)}{[k]}q^{\frac{k}{2}}w^k,&&
S_j^{-}(w) = \sum_{k=1}^{\infty}
\frac{a_j(k)}{[k]}q^{\frac{k}{2}}w^{-k},\\
a_{n-1}^*(k) = \sum_{l=1}^{n-1}\frac{-[lk]}{[k][nk]}a_l(k), &&
a_1^*(k) = \sum_{l=1}^{n-1}\frac{-[(n-l)k]}{[k][nk]}a_l(k).
\end{eqnarray*}

\end{document}